%% file: entcs.tex
\documentclass{entcs} \usepackage{entcsmacro}
\usepackage{amsmath}
\usepackage{upgreek}
\usepackage{color}

\usepackage{graphicx}
\usepackage{listings}
\usepackage{color}
\usepackage{caption}
\usepackage{epstopdf}
\usepackage{subfig}


\input{matt-might-macros}

\input{local-macros}

\mathchardef\mhyphen="2D

\definecolor{dkgreen}{rgb}{0,0.6,0}
\definecolor{gray}{rgb}{0.5,0.5,0.5}
\definecolor{mauve}{rgb}{0.58,0,0.82}

\def\lastname{Liang, Might, and Van Horn}
\begin{document}
\begin{frontmatter}
  \title{AnaDroid: Malware Analysis of  Android with User-supplied Predicates}
   \author{Shuying Liang%
  \thanksref{myemail},\thanksref{report}}
   \author{Matthew Might%
   \thanksref{myemail}}
     \address{School of Computing,  University of Utah\\
            Salt Lake City, Utah,  USA}  
            \author{David Van Horn%
            \thanksref{myemail}}
             \address{College of Computer and Information Sciences\\ Northeastern University\\ Boston, Massachusetts, USA}  
\thanks[myemail]{
  Supported by the DARPA Automated Program Analysis for Cybersecurity Program.
}
%
%
%
\thanks[report]{
 An extended report is available:
 \url{http://matt.might.net/a/2013/05/25/anadroid/}
}
\begin{abstract} 
 \input{abstract}

\end{abstract}
\end{frontmatter}
%
%
%
%
%


\input{content}
\bibliographystyle{acm}
\bibliography{mattmight,shuyingliang,local}
\vfill
\pagebreak
\vfill
\pagebreak

\end{document}

%% file: matt-might-macros.tex
\newcommand{\ie}{\emph{i.e.}}

\newcommand{\etal}{\emph{et al.}}

\newcommand{\syn}[1]{\mathsf{#1}}

\newcommand{\s}[1]{\mathit{#1}}

\newcommand{\parto}{\rightharpoonup}

\newcommand{\Pow}[1]{{\mathcal{P}\left(#1\right)}}

\newcommand{\To}{\mathrel{\Rightarrow}}

\newcommand{\wt}{\sqsubseteq}

\newcommand{\join}{\sqcup}

\newcommand{\bigjoin}{\bigsqcup}

\newcommand{\sembr}[1]{\ensuremath{[\![{#1}]\!]}}

\newcommand{\opor}{\mathrel{|}}

\newcommand{\produces}{\mathrel{::=}}

\newcommand{\stmt}{s}

\newcommand{\aexpr}{\mbox{\sl {\ae}}}

\newcommand{\store}{\sigma}

\newcommand{\addr}{a}

\newcommand{\aTo}{\leadsto}

\newcommand{\aInject}{{\hat{\mathcal{I}}}}

\newcommand{\sa}[1]{\widehat{\mathit{#1}}}

\newcommand{\aArgEval}{{\hat{\mathcal{A}}}}

\newcommand{\aStore}{\sa{Store}}

\newcommand{\astore}{{\hat{\sigma}}}

\newcommand{\acont}{{\hat{\kappa}}}

\newcommand{\aden}{{\hat{d}}}

\newcommand{\aaddr}{{\hat{\addr}}}


%% file: local-macros.tex
\newcommand{\cexpr}{\mathit{ce}}

\newcommand{\afp}{\hat{\mathit{fp}}}

\newcommand{\reg}{\mathit{name}}

\newcommand{\aconf}{{\hat c}}

\newcommand{\aFieldAddr}{\sa{FieldAddr}}

\newcommand{\afa}{\sa{fa}}

\newcommand{\aKontAddr}{\sa{KontAddr}}

\newcommand{\ka}{a_{\kappa}}
\newcommand{\aka}{\sa{\ka}}
\newcommand{\akainit}{\aka_{0}}

\newcommand{\aRegAddr}{\sa{RegAddr}}

\newcommand{\ara}{\sa{ra}}

\newcommand{\fp}{\s{fp}}

\newcommand{\aobjp}{\sa{op}}

\newcommand{\objv}{\mathit{ov}}
\newcommand{\aobjv}{\hat{\objv}}

\newcommand{\ObjectPointer}{\s{ObjectPointer}}
\newcommand{\aObjectPointer}{\widehat{\ObjectPointer}}

\newcommand{\FramePointer}{\s{FramePointer}}
\newcommand{\aFramePointer}{\widehat{\FramePointer}}

\newcommand{\FieldEval}{\mathcal{A}_{\mathcal{F}}}
\newcommand{\aFieldEval}{\hat\FieldEval}

\newcommand{\allocFP}{\s{allocFP}}
\newcommand{\allocOP}{\s{allocOP}}
\newcommand{\allocK}{\s{allocK}}

\newcommand{\aallocFP}{\widehat{\allocFP}}
\newcommand{\aallocOP}{\widehat{\allocOP}}
\newcommand{\aallocK}{\widehat{\allocK}}

\newcommand{\StmtOf}{{\mathcal{S}}}



%% file: abstract.tex
Today's mobile platforms provide only coarse-grained permissions to users with
regard to how third-party applications use sensitive private data.
Unfortunately, it is easy to disguise malware within the boundaries of
legitimately-granted permissions.
For instance, granting access to ``contacts'' and ``internet'' may be necessary
for a text-messaging application to function, even though the user does not
  want contacts transmitted over the internet.
To understand fine-grained application use of permissions, we need to statically analyze
their behavior.
%
%
Even then, malware detection faces three hurdles: (1) analyses may be
prohibitively expensive, (2) automated analyses can only find behaviors that
they are designed to find, and (3) the maliciousness of any given behavior is
application-dependent and subject to human judgment.
%
%
To remedy these issues,  we propose 
semantic-based program
analysis, with a human in the loop
as an alternative approach to malware detection.
In particular, our analysis allows analyst-crafted semantic predicates to search and filter analysis results. 
Human-oriented semantic-based program analysis can 
systematically, quickly and concisely characterize the behaviors of mobile applications.
We describe a tool that provides analysts with a library of the semantic predicates and
the ability to dynamically trade speed and precision.
It also provides analysts the ability to statically inspect details
of every suspicious state of (abstract) execution in order to make a ruling
as to whether or not the behavior is truly malicious with respect to
the intent of the application.
In addition, permission and profiling reports are generated to
aid analysts in identifying common malicious behaviors. 
%
%

%% file: content.tex
\section{Introduction}

Google's Android is the most popular mobile platform, with a share of 52.5\% of all smartphones~\cite{local:Gartner:url}.
Due to Android's open application development community, more than 400,000 apps are available 
with 10 billion cumulative downloads by the end of 2011~\cite{local:google10b:url}.

While most of those third-party  applications have  legitimate reasons to access private data, 
the grantable permissions are too coarse: malware can hide in the cracks.
For instance, an app that should only be able to read information from a specific site and have access to GPS information
must necessarily be granted full read/write access to the entire internet, thereby allowing a possible location leak over the net.
%
Or, a note-taking application can wipe out SD card files when a hidden trigger condition is met.
Meanwhile, a task manager that requests every possible permission
can be legitimately benign.

To understand fine-grained use of security-critical resources, we need to statically analyze the application
with respect to what data is accessed, where the sensitive data flows, and what operations have been performed 
on the data (\ie, determine whether the data is tampered with).  
Even then, automated malware detection faces three hurdles: (1) analyses may be
prohibitively expensive, (2) automated analyses can only find behaviors that
they are designed to find, and (3) the maliciousness of any given behavior is
application-dependent and subject to human judgment.

In this work, we propose 
semantics-based program
analysis with a human in the loop
as an alternative approach to malware detection.
Specifically, we derive an analytic engine, an abstract CESK* machine based on   
the design methodology of Abstracting Abstract Machines (AAM)~\cite{VanHorn:2010:Abstract} 
  to analyze object-oriented bytecode. 
Then we extend the foundational analysis to analyze specific  features:  multiple entry points of Android apps and reflection APIs.
Finally, 
we  describe a tool that provides
analysts with a library of  semantic predicates that can be used to search and filter analysis results, 
and the ability to dynamically trade speed and precision.
The tool also provides analysts the ability to statically inspect details
of every suspicious state of (abstract) execution in order to make a ruling
as to whether or not the behavior is truly malicious with respect to
the intent of the application.
%
%
Human-oriented, semantics-based program analysis can 
systematically characterize the behaviors of mobile applications.

\paragraph{Overview}
The remainder of the paper is organized as follows: 
Section~\ref{sec:analysis} presents the syntax of an object-oriented byte code, and  
 illustrates a finite-state-space-based 
 abstract interpretation of the byte code.
 Section~\ref{sec:analysis-specifics} discusses analysis techniques to analyze Android-specific issues: multiple entry points and reflection APIs. 
 Section~\ref{sec:tool} presents the tool implementation  with user-supplied predicates. 
Section~\ref{sec:related} discusses related work, and
 Section~\ref{sec:conclusion} concludes.

 \section{Semantic-based program analysis}\label{sec:analysis}
 
Android apps are written in Java, and 
compiled into Dalvik virtual machine byte code (essentially a register-based version of Java byte code).
In this section,  we  present  how to derive an analysis for  
 a core object-oriented (OO) byte code   language based on Dalvik.
After presenting this foundational analysis, 
  we shall illustrate Android-specific analysis techniques 
  in subsequent sections.

\input{oo-syntax}
   The first step 
    is to define a syntax. 
Figure~\ref{fig:oo-syntax} presents the  syntax of  an OO byte code language 
that is  closely modeled on the Dalvik virtual machine.
   Statements encode individual actions; 
     atomic expressions encode atomically computable values;
     and complex expressions encode expressions with possible 
     non-termination or side effects.
     There are four kinds of names:
     $\syn{Reg}$ for registers, $\syn{ClassName}$ for class names,
     $\syn{FieldName}$ for field names
     and $\syn{MethodName}$ for method names.
     The special register name
     $\syn{ret}$ holds the return value of the last function called.
     With respect to a given program, we assume a syntactic meta function
     $\StmtOf : \syn{Label} \to \syn{Stmt}^*$, which maps a label to the
     sequence of statements that start with that label.

  Ordinarily, the next step toward an analyzer would be to derive a concrete machine to interpret  the  language just defined.
  The meaning of a program will be defined as the set of machine states reachable from an initial state.
  The purpose of static analysis is to derive a computable approximation of
  the concrete machine's behavior---of these states.
  We'll construct an abstract semantics to do that.
  Since the concrete and the abstract semantics are so close in structure,
  we will present only the abstract semantics of the byte code,  
  while highlighting places that are different from its concrete counterpart to save space.

\input{sec-abstract-semantic.tex}

\section{Analysis of reflection and  multiple entry points in abstract CESK* machine \label{sec:analysis-specifics}}
 
While the crux of the  analysis for Android apps has been presented in the previous section, 
our abstract CESK* machine has to be extended to analyze multiple entry points in Android apps and  some intricate APIs like reflection.

\subsection{Fixed-point computation for multiple entry points\label{sec:fep}}
A typical static analysis only deals with one entry point of traditional programs (the \textbf{main} method).
However, any Android application has more than one entry point, due to the event-driven nature of the Android platform.
Intuitively, to explore the reachable states for all the entry points seems to require the exploration for all the permutation of  entry points.
But this can easily lead to state-space explosion. 
Related works like~\cite{shuyingliang:Lu:2012:CCS:CHEX} prune paths for specific Android apps (but not soundly).
We solve the problem in a sound but inexpensive way. 
Specifically, we iterate over all entry points that have been found.
For each entry point, we compute its reachable states via the abstract CESK* machine.
Then we compute a single widened store from those states using the widening techniques similar to the ones presented in~\cite{Might:2007:Dissertation}. 
The store  then forms  part of next state to continue the next entry point fixed-point computation. 
Obviously, the store is monotonic, which  ensures a sound approximation of the effects introduced from all entry points.
This diminishes precision slightly, but the gains in speed are considerable.
The  effects of similar technique are also noted in~\cite{Might:2007:Dissertation}\cite{mattmight:Shivers:1991:CFA}.
 
\input{sec-reflection.tex}

\section{The tool with user-supplied predicates \label{sec:tool}}  

In this section, we first briefly presents the implemented analyzer, since the core of the analysis has been specified in Section~\ref{sec:analysis} an~\ref{sec:analysis-specifics}.
Then we illustrate the tool usage, particularly with respect to user-supplied predicates.

The analysis engine is a faithful rendering of  the formal specification in Section~\ref{sec:analysis} and~\ref{sec:analysis-specifics}.
In addition, it incorporates previous techniques that  boost precision and performance, 
including the abstract garbage collection~\cite{Might:2006:GammaCFA}, 
store-widening~\cite{mattmight:Might:2007:Dissertation}, 
and simple abstract domains to analyze strings~\cite{shuyingliang:Costantini:2011:String}.
Other constructs of the tool are:
\begin{itemize}
\item{\textbf{Entry points finder:} \label{subsec: multi-ens}}
This component  discovers all the entry points of an Android application. Then the engine  will explore reachable states based on the algorithm presented in Section~\ref{sec:analysis-specifics}.
 
\item {\textbf{Permission violation report:}}
It  reports whether an application
asks for more permissions than it actually uses or vice versa.

\item\textbf{State graph:} This presents all reachable states with states-of-interest  highlighted
according to default predicates  or those supplied by analysts.
\item \textbf{{API dumps:}} This presents all the reachable API calls.
\item\textbf{{Heat map:}} This reports analyzer intensity on per-statement basis.
\end{itemize}
 
 
  The work flow of our human-in-the-loop analyzer---AnaDroid---is as follows:
 (1) an analyst configures analysis options and malware predicates;
 (2) AnaDroid presents a permission-usage report, an API call dump, a state graph and a heat map.  
 The major parameters of the analyzer
 include call-site context-sensitivity---$k$,    
 widening,
 abstract garbage collection, 
 cutoffs,
 and predicates.
 An analyst can make  the trade-off between runtime and precision of the analyzer with these parameters.
 In addition, the predicates enable  analysts to inspect states of interests to detect malware.

 \subsection{\textbf{Semantic predicates}}
 To assist analysts,  we provide a library of  predicates
  for common  patterns.  The two major kinds of predicates in AnaDroid are:
 ``State color predicate''  renders matching states in a customized color;
   ``State truncate predicate'' optimizes the analysis exploration  
   by allowing analysts to manually prune paths beginning at matching states.
 Examples of usage of the predicates are listed as follows:

 
  {\tt{uses-API?:}} It is used to specify what color to render the state that uses the specified API call.
 The color is a string representing a SVG color scheme~\cite{local:svgcolor:url}, \ie``red, colorscheme=set312'':

\begin{lstlisting}[frame=single, language=Lisp,numbers=none,  belowcaptionskip=0em,
    belowskip=0em]   
 (lambda (state)
   (if (uses-API? state "org/apache/http/client/HttpClient/execute" st-attr)
       "red,colorscheme=set312"
       #f))
      \end{lstlisting} 
 Note that {\tt{st-attr}} is a specialized keyword used by our analyzer. {\tt{state}} is the parameter of the predicate.
 
 {\tt{uses-name?:}} It is a variant of  {\tt{uses-API?}}, used to identify a state with the  specific method name in code:
\begin{lstlisting}[frame=single, language=Lisp,numbers=none,  belowcaptionskip=0em,
    belowskip=0em]   
 (lambda (state)
   (if (uses-name?  state "org/ucomb/android/testinterface/RectanglePlus/getArea")
       "red,colorscheme=set312"
       #f))
    \end{lstlisting}
 
  An analyst can also use {\tt{cond}}
  to specify multiple colors:
  
 
\begin{lstlisting}[frame=single, language=Lisp,numbers=none,  belowcaptionskip=0em,
    belowskip=0em]   
 (lambda (state) 
   (cond
     [(uses-API? state "org/apache/http/client/HttpClient/execute" st-attr )  "red,colorscheme=set312"]
     [(uses-name? state  "org/ucomb/android/testinterface /RectanglePlus/getArea") "8,colorscheme=set312"]
     [else #f]))
   \end{lstlisting}
 
 {\tt{truncate?}}: 
 
 \begin{lstlisting}[frame=single, language=Lisp,numbers=none,  belowcaptionskip=0em,
     belowskip=0em]   
 (lambda (state)
   (if (truncate? state "org/apache/http/client/HttpClient/execute")
       "12,colorscheme=set312"
       #f)) 
      \end{lstlisting}

 %
 %
 %
 %

     \begin{figure*}
     \centering
 
             
 
     \end{figure*}



\section{Related work} \label{sec:related}

Stowaway~\cite{shuyingliang:Felt:2011:Stowaway} is a static analysis tool 
identifying whether an application requests more permissions than it actually uses. 
PScout~\cite{shuyingliang:Au:2012:CCS:PScout} aims for a similar goal, but produces more precise and  fine-grained mapping from APIs to permissions.
Our least permission report uses the Stowaway permission map as AnaDroid's database.\footnote{Our new version analyzer is upgraded to  PScout data set.}
%
%
%


Jeon~\etal{}~\cite{shuyingliang:Jeon:2012:DAM} proposes enforcing a fine-grained permission system. It limits access to resources 
 that could  be accessed by  Android's default permissions.
 Specifically, the security policy uses a white list to determine which resources an app can use and a black list to deny access to resources.
 In addition,  strings potentially containing URLs are  identified by pattern matching and constant propagation is used to infer more specific Internet permissions.

Dynamic taint analysis has been applied to identify security vulnerabilities at run time in Android apps.
 TaintDroid~\cite{shuyingliang:Enck:2010:TaintDroid} dynamically tracks the flow of sensitive information and looks for confidentiality violations.
 QUIRE~\cite{shuying:Dietz:2011:Quire}, IPCInspection~\cite{shuyingliang:Felt:2011:IPCInspection}, and XManDroid~\cite{shuyingliang:Bugiel:2012:XManDroid}  are designed to prevent privilege escalation, 
 where an application is compromised  to provide sensitive capabilities to other applications. 
 The vulnerabilities introduced by interapp communication is considered future work.
 However, these approaches typically ignore implicit flows raised by control structures in order to reduce run-time overhead. 
 
The other approach to enforce security on mobile devices is  delegating the control to users.
iOS and Window User Account Control~\cite{local:MS:url} can prompt a dialog to request  permissions from 
users when applications try to access resources  or make security  or privacy-related system level  changes. 
However,  we advocate stopping potential malware from floating in the market beforehand via strict  inspections.
Our tool has designed with analysts in mind and can help them identify 
malicious behaviors of submitted applications.

\section{Conclusion\label{sec:conclusion}}


In this work, we propose a human-oriented semantic-based program analysis for Android apps.
We derive an abstract CESK* machine to analyze  object-oriented bytecode.
Then the foundational analysis is extended 
  to analyze specific  features:  multiple entry points of Android apps and reflection APIs.
We also  describe a tool that provides
analysts with a library of  semantic predicates that can be used to search and filter analysis results, 
and the ability to dynamically trade speed and precision.
It also provides analysts the ability to statically inspect details
of every suspicious state of (abstract) execution in order to make a ruling
as to whether or not the behavior is truly malicious with respect to
the intent of the application.
In addition, permission and profiling reports are generated to
aid analysts in identifying common malicious behaviors. 
The technique can systematically, quickly and concisely characterize the behaviors of mobile applications,
as demonstrated by case studies in the extended report.
%
%


%% file: oo-syntax.tex
\begin{figure} 
\footnotesize{
 \begin{align*}
\mathit{program} &\produces 
\mathit{class} \mhyphen \mathit{def}\; \ldots
\\
\mathit{class} \mhyphen \mathit{def} \in \syn{ClassDef}&\produces (\mathit{attribute}\;\ldots~ \syn{class} ~\mathit{class}\mhyphen\mathit{name} ~\syn{extends}  ~\mathit{class}\mhyphen\mathit{name} 
\\
&\;\;\;\;\;\;\;\; (\mathit{field}\mhyphen{\mathit{def}}  \dots)\; (\mathit{method}\mhyphen\mathit{def} \dots))
\\
\mathit{field}\mhyphen\mathit{def} &\produces  (\syn{field}~\mathit{attribute}\;\ldots~\mathit{field}\mhyphen\mathit{name}~\mathit{type})
\\
\mathit{method}\mhyphen\mathit{def} \in \syn{MethodDef} &\produces  (\syn{method}~\mathit{attribute}\;\ldots~\mathit{method}\mhyphen\mathit{name}~(\mathit{type} \dots)~\mathit{type} 
\\
&\;\;\;\;\;\;\;\;
(\syn{throws}~\mathit{class}\mhyphen\mathit{name} \dots)
\;
(\syn{limit} ~n)
\; 
\stmt\; \ldots)
\\
\stmt \in \syn{Stmt}&\produces (\syn{label} ~\mathit{label}) \opor (\syn{nop}) \opor (\syn{line} ~\mathit{int}) \opor  (\syn{goto} ~\mathit{label}) 
\\
&\;\;\opor\;\;  (\syn{if}\; \aexpr ~(\syn{goto}~\mathit{label})) \opor (\syn{assign}~\mathit{\reg}\; [\aexpr \opor \cexpr]) \opor (\syn{return} ~\aexpr) 
\\
&\;\;\opor\;\; (\syn{field}\mhyphen\syn{put} ~\aexpr_o ~\mathit{field}\mhyphen\mathit{name} ~\aexpr_v) \opor (\syn{field}\mhyphen\syn{get}\; \reg ~\aexpr_o ~\mathit{field}\mhyphen\mathit{name})
\\
\aexpr \in \syn{AExp} &\produces \syn{this} \opor \syn{true} \opor \syn{false} \opor \syn{null} \opor \syn{void} \opor \mathit{\reg} \opor \mathit{int} 
\\
&\;\;\opor\;\;  (\mathit{atomic}\mhyphen\mathit{op} ~\aexpr \dots \aexpr)  \opor \syn{instance}\mhyphen\syn{of}(\aexpr, \mathit{class}\mhyphen\mathit{name}) 
\\
\cexpr &\produces (\syn{new}\; \mathit{class}\mhyphen\mathit{name})  
\\
&\;\;\opor\;\;(\mathit{invoke}\mhyphen\mathit{kind} \; \mathit{method}\mhyphen\mathit{name}~(\aexpr\dots\aexpr)\; (\mathit{type}_0\;\dots\;\mathit{type}_n))
\\
\mathit{invoke}\mhyphen\mathit{kind} &\produces \syn{invoke}\mhyphen\syn{static}  \opor \syn{invoke}\mhyphen\syn{direct} \opor \syn{invoke}\mhyphen\syn{virtual} 
\\
&\;\;\opor\;\;\syn{invoke}\mhyphen\syn{interafce}  \opor \syn{invoke}\mhyphen\syn{super}
\\
\mathit{type} &\produces ~\mathit{class}\mhyphen\mathit{name} \opor \syn{int} \opor \syn{byte} \opor \syn{char} \opor \syn{boolean}  å
\\
\mathit{attribute} &\produces \syn{public} \opor \syn{private} \opor \syn{protected} \opor \syn{final} \opor \syn{abstract} 
\text.
\end{align*}
}
\caption{An object-oriented bytecode adapted from the Android specification~\cite{local:androidbytecode:url}.}


\label{fig:oo-syntax}
\end{figure}

%% file: sec-abstract-semantic.tex
\subsection{Abstract semantics \label{sec:abstract-semantics}}

We define our abstract interpretation as a direct, structural
abstraction of a machine model for the OO bytecode~\cite{VanHorn:2010:Abstract}.  
Because the structural abstraction creates an abstract machine nearly identical to the
machine model itself (with exceptions that we explain),
we don't provide the concrete machine model.
The
analysis of a program is defined as the set of \emph{abstract}
machine states reachable by an abstract transition relation $(\aTo)$---the core of the abstract semantics.
That is, abstract evaluation is defined by the set of states reached by
the reflexive, transitive closure of the $(\aTo)$ relation.


Figure~\ref{fig:abs-conf-space} details the abstract state-space. 
We assume the natural element-wise, point-wise and member-wise
lifting of a partial order $(\wt)$ across this state-space.
 States of this machine consist of a tuples of
     of statements, frame pointers, heaps, and stack pointers.
To synthesize the abstract state-space,
we force frame pointers and object pointers (and thus addresses) to be a finite set.
When we compact the set of addresses into a finite set during a structural abstraction, 
the machine may (and likely will) run out of addresses to allocate, and when it does, 
the pigeon-hole principle will force multiple abstract values to reside at the same (now abstract) address. 
As a result, we have no choice but to force the range of the $\aStore$ to become a power set 
in the abstract state-space: now each abstract address can hold multiple values.

\input{abstract-state-space}

\subsubsection{Abstract transition relation} \label{sec:abs-transition-rules}

In this section, we provide major cases for the abstract transition relation.
%
The abstract transition relation delegates 
to helper functions 
for injecting programs into states,
and for evaluating atomic
expressions and looking up field values:

\begin{itemize}

\item
$\aInject : \syn{Stmt^*} \to \sa{State}$ injects an sequence of instructions into
                                              an initial state:
                                              $
\aconf_0 = \aInject(\vec{s}) = (\vec{s}, \afp_0, [ \akainit \mapsto \syn{halt}], \akainit)
$

\item
$\aArgEval : \syn{AExp} \times \aFramePointer \times \aStore \rightharpoonup
  \sa{Val}$ evaluates atomic expressions (specifically for variable look-up):
  $   \aArgEval(\reg,\afp,\astore) = \store(\afp,\reg) 
 $
\item
$ \aFieldEval : \syn{AExp} \times  \aFramePointer \times \aStore \times \syn{FieldName} \rightharpoonup \sa{Val}$
looks up fields:
{\fontsize{10.5pt}{10.5pt}\selectfont
  \begin{align*}
    \aFieldEval(\aexpr_o,\afp,\astore, \mathit{field}\mhyphen \mathit{name}) &=  \bigjoin \astore(\aobjp,\mathit{field}\mhyphen \mathit{name})  ~\text{, where}
    \\
    ~(\aobjp, \mathit{class}\mhyphen \mathit{name}) &\in  \aArgEval(\aexpr_o, \afp, \astore )
    \text.
    \end{align*}
}

\end{itemize}


 The rules for the abstract transition relation $(\aTo) \subseteq \sa{State} \times
    \sa{State}$ describe how components of state evolve in light
    of each kind of statement.
In subsequent paragraphs, we will illustrate the important rules that
involve objects and function calls, omitting less important ones to save space:

%
 
 %

\begin{itemize}

\item{ {\textit{New object creation}}} 
Creating a new object allocates a potentially non-fresh address and joins the
newly initialized object to other values residing at this store address.
{\fontsize{10.5pt}{10.5pt}\selectfont
\begin{align*}
  \overbrace{
 (\sembr{(\syn{assign}\; \reg\; (\syn{new}\;\mathit{class}\mhyphen\mathit{name})) :  \vec{s}}, \afp, \astore,  \aka)}^{\aconf}
    \To 
    (\vec{s}, \afp, \astore'',  \aka), \text{where}
   \\
  \aobjp' = \aallocOP(\aconf), \;
   \astore' = \astore \join [(\afp, \mathit{\reg}) \mapsto  ( \aobjp', \mathit{class}\mhyphen\mathit{name})], 
  \\
  \astore'' = \widehat{\mathit{initObject}}(\astore', \mathit{class}\mhyphen\mathit{name})
  \text{,} 
    \end{align*} 
    }
where the  helper 
{\fontsize{10.5pt}{10.5pt}\selectfont
$\widehat{\mathit{initObject}}: \aStore \times \syn{ClassName} \rightharpoonup \aStore$ } initializes fields. 
  
  \item{\textit{Instance field reference/update}}
Referencing a field uses $\aFieldEval$ to lookup the field values and joins these values with the values at the store location for the destination register:
{\fontsize{10.5pt}{10.5pt}\selectfont
 \begin{align*}
     (\sembr{(\syn{field}\mhyphen\syn{get}\; \reg ~\aexpr_o ~\mathit{field}\mhyphen\mathit{name}) : \vec{s}}, \afp, \astore,\aka)
   &\aTo
     (\vec{s},\afp,\astore', \aka)
  \text{, where}
  \\
  \astore' = \astore \join [ (\afp, \reg)&\mapsto \aFieldEval(\aexpr_o, \afp, \astore, \mathit{field}\mhyphen\mathit{name} ) ] 
  \text.
   \end{align*}
   }
   
   Updating a field first determines the abstract object values from the store,
   extracts the object pointer from all the possible values,
   then pairs the object pointers with the field name to get the field address,
   and finally \textit{joins} the new values to those found at this store location:
   
{\fontsize{10.5pt}{10.5pt}\selectfont
 \begin{align*}
     (\sembr{(\syn{field}\mhyphen\syn{put}~\aexpr_o~\mathit{field}\mhyphen{name} ~\aexpr_v) : \vec{s}}, \afp, \astore, \aka)
   \aTo
     (\vec{s},\afp,\astore', \aka)
  \text{, where}
  \\
  \astore' = \astore \join [(\aobjp, \mathit{field}\mhyphen{name}) \mapsto \aArgEval(\aexpr_v, \afp, \astore)], \;
  (\aobjp, \mathit{class}\mhyphen{name}) &\in \aArgEval(\aexpr_o, \afp, \astore)\text.
   \end{align*}
   }

\item{{\textit{Method invocation}}}
This rule involves all four components of the machine.
The abstract interpretation of non-static method invocation can result in the
method being invoked on a \emph{set} of possible objects, rather than a single
object as in the concrete evaluation.
Since multiple objects are involved, this can result in different method definitions being resolved for the different objects.
The method is resolved\footnote{Since the language supports inheritance, method
resolution requires a traversal of the class hierarchy.  This traversal follows
the expected method and is omitted here so we can focus  on the abstract
rules.} and then applied as follows: 
{\fontsize{10.5pt}{10.5pt}\selectfont
$  \overbrace{
    (\sembr{ (\mathit{invoke}\mhyphen\mathit{kind}\; \mathit{method}\mhyphen\mathit{name} ~(\aexpr_0\dots\aexpr_n)\; (\mathit{type}_0\dots\mathit{type}_n))} :  \vec{s}, \afp, \astore, \aka)
    }^{\aconf}$
    
  $\;\;\;\;\;\;\;\;\;\;\;\;\;\;\;\;\;\;\;\;\;\;\;\;\;\;\;\;\;\;\;\;\;\;\;\;\;\;\;\;\;\;\;\;\;\;\;\;\;\;\;\;\;\;\;\;\;\;\;\; \aTo 
     \widehat{\mathit{applyMethod}}(m, \vec{\aexpr},
     \afp,\astore,\aka)$}

where the  function ${\widehat{\mathit{applyMethod}}}$ takes a method definition,
arguments,
a frame pointer,
a store, and a new stack pointer  and produces the next states:
{\fontsize{10.5pt}{10.5pt}\selectfont
\begin{align*}
 \widehat{\mathit{applyMethod}}(m, \vec{\aexpr}, \afp, \astore, \aka) &= (\vec{\stmt}, \afp', \astore'', \aka'), \text{where}
\\
\afp' =  \aallocFP(\aconf), &~~~~ \aka' = \aallocK(\aconf),
\\
\astore' = \astore \join [\aka' \mapsto \{\mathbf{fun}(\afp, \vec{s}, \aka)\}] ,&~~~~\astore'' = \astore' \join [(\afp', \reg_i) \mapsto \aArgEval(\aexpr_i, \afp, \astore)]\text.
\end{align*}}

\item{\textit{Procedure return}}
Procedure return restores the caller's context and \textit{extends} the return value in
the dedicated return register, $\syn{ret}$.
{\fontsize{10.5pt}{10.5pt}\selectfont
\begin{align*}
      (\sembr{(\syn{return}~\aexpr) :  \vec{s}}, \afp, \astore,  \aka)
    &\aTo 
       (\vec{s}', \afp', \astore',  \aka')
   \text{, where}
      \\
     \mathbf{fun}(\fp',\vec{s'}, \ka') \in \store(\ka) &~\text{and}~ \astore' = \astore\join [(\afp', \syn{ret})\mapsto \aArgEval(\aexpr, \afp, \astore)]
   \text.
    \end{align*}}
\end{itemize}

%% file: abstract-state-space.tex
 \begin{figure}
 \footnotesize{
\begin{align*}
 \aconf \in \sa{State} &= \syn{Stmt^*} \times \sa{FramePointer} \times \sa{Store} \times \aKontAddr && \text{[states]}
\\
 \astore \in \sa{Store} &= \sa{Addr}  \parto   \sa{Val} && \text{[stores]}
 \\
\aaddr \in  \sa{Addr} &=  \aRegAddr  + \aFieldAddr   + \aKontAddr
&& \text{[addresses]}
 \\
 \aka \in \aKontAddr &\text{ is a finite set of continuation addresses}
 \\
  \ara \in \aRegAddr &= \sa{FramePointer}  \times \syn{Reg}  
  \\
 \afa \in \aFieldAddr &= \sa{ObjectPointer} \times {\syn{FieldName} }
\\
\acont \in \sa{Kont} &=  \mathbf{fun}(\afp,\vec{s},\aka) + \mathbf{halt}
&& \text{[continuations]}
\\
\aden \in \sa{Val} &= \Pow{ \sa{ObjectValue} +  \sa{String} +  \sa{\mathcal{Z}}+ \sa{Kont}} 
&& \text{[abstract values]}  
 \\
 \aobjv \in \sa{ObjectValue} &= \aObjectPointer \times \syn{ClassName}
\\
\afp \in \aFramePointer &\text{ is a finite set of frame pointers} && \text{[frame pointers]}
\\ 
\aobjp \in \aObjectPointer&\text{ is a finite set of object pointers} && \text{[object pointers]}
\text.
 \end{align*}}
\caption{The abstract state-space.}
 \label{fig:abs-conf-space}
 \end{figure}

%% file: sec-reflection.tex
\subsection{Reflection}

This section presents  how to extend the abstract CESK* machine to analyze one of the most commonly used dynamic features in Android---reflection.

Reflection enables programs to access class information to create objects and invoke methods at runtime. Type information involved is dynamically retrieved  from strings. The strings can come from user input, files, network or hard-coded, literal strings.  Literal strings are not infrequent in reflection. The following code snippet demonstrates a common case of reflection in Java: 

\begin{lstlisting}[frame=single,  belowcaptionskip=0em,
    belowskip=0em]   
 Class<?> aeco = Class.forName("android.os.Environment");
 Method externalDir = aeco.getMethod("getExternalStorageDirectory", (Class[])null);
 (File)externalDir.invoke(null); 
\end{lstlisting}
A class object is created in Ln.1 and the method object for \textit{getExternalStorageDirectory}\footnote{Since the method is a static method, so no instantiated object is needed, which is $\syn{null}$.} is created in Ln.2. Finally, the method is invoked  in Ln.3 via the method object  \textit{externalDir}. Since it is a static method with no arguments, the receiver object being invoked  is $\syn{null}$. Otherwise, the argument $\mathit{aeco.newInstance()}$  needs to be supplied  in  Ln.3.
 
To analyze such reflection,  we can integrate string analysis into abstract interpretation of Java API calls.
In the abstract CESK*, we need five  additional transition rules, 
mainly for simple string analysis and the APIs involving   creation of class object, method object, class instantiation and method invocation:

\begin{itemize}
\item {\textit{String:}}
Strings are objects in Java, and so string instantiation is a special case for the $\syn{new}$ rule 
(see Section~\ref{sec:abstract-semantics})\footnote{java/lang/StringBuilder is interpreted in similar way.}:

{\fontsize{10.5pt}{10.5pt}\selectfont
$ \overbrace{
 (\sembr{( \syn{const} \mhyphen\syn{string}  ~\mathit{name} ~ \mathit{str}) :  \vec{s}}, \afp, \astore,  \aka)}^{\aconf}
    \aTo 
    (\vec{s}, \afp, \astore'',  \aka), \text{where}\; \aobjp = \aallocOP(\aconf).$
    
 $  \astore' = \astore \join [(\afp, \mathit{\reg})\mapsto\{( \aobjp, \syn{java/lang/String)}\}], \; \astore'' =  \astore' \join [(\aobjp, \syn{value}) \mapsto \alpha(\mathit{str})],$
    }
     Unlike the usual case for $\syn{new}$ rule,   there is  a field $\syn{value}$ paired with the string object pointer as field offset
    to store abstract string values. $\alpha$ is the abstraction function for string values. The simplest form is to construct a flat lattice for strings. 
    Other string analysis such as Costantini~\etal{}~\cite{shuyingliang:Costantini:2011:String}, Christensen~\etal{}~\cite{shuyingliang:Christensen:PreciseStringAnalysis:2003}, \emph{etc{.}} can be directly incorporated.

\item{\textit{Class objects:}}
In byte code, the creation of a class object using $\syn{Class.forName}$ is an $\syn{invoke}\mhyphen\syn{static}$ statement, with the first argument referencing to string values. The rule will allocate a new class object on the heap, with the field offset $\syn{class}\mhyphen\syn{name}$ points to the  string reference looked up by the address $(\afp, \aexpr)$. In addition, the class object reference is stored into the $\syn{ret}$ address:

{\fontsize{10.5pt}{10.5pt}\selectfont
 $ \overbrace{
 (\sembr{( \syn{invoke} \mhyphen\syn{static}   \;\syn{java/lang/Class/forName} \; \aexpr \; \syn{java/lang/String}) :  \vec{s}}, \afp, \astore,  \aka)}^{\aconf}$
  
  $\;\;\; \aTo 
     (\vec{s}, \afp, \astore'',  \aka), \text{where} \;  \aobjp_\syn{Cls} = \aallocOP(\aconf)$,
 }
 
 {\fontsize{10.5pt}{10.5pt}\selectfont
$  \astore' =  \astore  \join [(\aobjp_\syn{Cls}, \syn{class}\mhyphen\syn{name}) \mapsto  \astore(\afp, \ae) ]$, 
  
 $    \astore'' = \astore' \join [(\afp, \syn{ret}) \mapsto  ( \aobjp_\syn{Cls}, \syn{java/lang/Class)}] $
    }

\item{\textit{Method objects:}}
Method objects are represented as  method headers, including function name, arguments and their types, return values and exceptions that the method can  throw.\footnote{Exceptions handling is omitted in the semantics.}
A method object is resolved from a class object, whose class name can be obtained from the first argument $\aexpr_0$. The second argument will be resolved as the method name. Arrays of argument  types of the method object are stored in the third register $\aexpr_3$.\footnote{We don't interpret the arrays of the types explicitly.}

{\fontsize{10.5pt}{10.5pt}\selectfont
 $ \overbrace{
 (\sembr{( \syn{invoke} \mhyphen\syn{virtual} \; \syn{java/lang/Class/getMethod} \;(\aexpr_0 \;\aexpr_1\; \aexpr_2)\; \mathit{types_{\mathit{args}}}) :  \vec{s}}, \afp, \astore,  \aka)}^{\aconf}$

\;\; $    \aTo 
   \mathit{newMethodObject}( \aobjp_\syn{Method}, \vec{m}, \vec{s}, \afp, \astore', \aka)  \text{, where}$
 }
 
 {\fontsize{10.5pt}{10.5pt}\selectfont
   \begin{align*}
   (\aobjp_{\syn{Cls}}, \syn{java/lang/Class}) \in   \astore(\afp, \aexpr_0),  \;
   (\aobjp_0, \syn{java/lang/String} ) &\in \astore(\aobjp_{\syn{Cls}}, \syn{class}\mhyphen\syn{name}), 
   \\ 
   \mathit{class}\mhyphen\mathit{name} \in \astore(\aobjp_0, \syn{value}), \;
    (\aobjp_1, \syn{java/lang/String} ) &\in \astore(\afp, \aexpr_1)   
      \\ 
      \mathit{method}\mhyphen\mathit{name} \in \astore(\aobjp_1, \syn{value}),\;
      \aobjp_\syn{Method} &= \aallocOP(\aconf) 
       \\ 
         \astore' = \astore \join [(\afp,  \syn{ret}) \mapsto  (\aobjp_\syn{Method}, \syn{java/lang/Reflect/Method)}]  
  \text{.} 
    \end{align*} 
    }
    Similarly like the transition rule for function call, the method resolution process is omitted here. The resolution process needs the information $\syn{class}\mhyphen\syn{name}$ and $\syn{method}\mhyphen\syn{name}$.  Also note that the resolution result is a set of public methods $\vec{m}$, rather than one.
    The helper function $newMethodObject$ takes the newly allocated method object pointer, the set of method definitions in the domain $\syn{MethodDef}$, the rest statements, the frame pointer, store, and the stack pointer and returns the successor states. Again, the method object value will be stored into the $\syn{ret}$ address.
    
\item{\textit{Class instantiation:}}
 The API call $\syn{java/lang/Class/newInstance}$ is used to instantiate a new object of a concrete class type (not an abstract class nor interface).
The class type name can be resolved from the first argument $\aexpr$ of the instruction.
Unlike the normal $\syn{new}$ statement, the class instantiation requires the invocation of default class constructor. Therefore, we first resolve class definitions by using  a helper function $\mathcal{C}: \Pow{\syn{ClassName}} \rightarrow \Pow{\syn{ClassDef}} $, and then use $\syn{getDefaultConstructor}: \syn{ClassDef}\rightarrow \syn{MethodDef}$ to get the a constructor method. After that,  the control is transferred to the constructor invocation via $\syn{invoke}\mhyphen\syn{direct}$ statement, which is inserted in front of the rest states $\vec{s}$.
 
{\fontsize{10.5pt}{10.5pt}\selectfont
 $ \overbrace{
 (\sembr{( \syn{invoke} \mhyphen\syn{virtual} \; \syn{java/lang/Class/newInstance} \;\aexpr \; \syn{java/lang/Class}) :  \vec{s}}, \afp, \astore,  \aka)}^{\aconf}$  }
  {\fontsize{10.5pt}{10.5pt}\selectfont
  
 $ \;\;\; \aTo 
      (s' : \vec{s}, \afp, \astore'',  \aka),  \text{where} $
\begin{align*}
   (\aobjp_{\syn{Cls}}, \syn{java/lang/Class}) \in   \astore(\afp, \aexpr), \;
   (\aobjp, \syn{java/lang/String} ) \in \astore(\aobjp_{\syn{Cls}}, \syn{class}\mhyphen\syn{name}),
   \\ 
   \mathit{class}\mhyphen \mathit{def} \in \mathcal{C}( \astore(\aobjp, \syn{value})), \;
      \aobjp_\syn{Cls} = \aallocOP(\aconf),
   \\
   \mathit{m} = \syn{getDefaultConstructor}(\mathit{class}\mhyphen\mathit{def}),
   \\
   {s}' = (\syn{invoke} \mhyphen\syn{direct},  \;m.\mathit{method}\mhyphen\mathit{name} ~(\aexpr_0\dots\aexpr_n)\; (\mathit{type}_0\dots\mathit{type}_n)),
   \\
     \astore' = \astore \join [(\afp,  \ae_{0}) \mapsto  (\aobjp_\syn{Cls}, \syn{java/lang/Class)}],
     \\
          \astore''= \astore' \join [(\afp, \syn{ret}) \mapsto  (\aobjp_\syn{Cls}, \syn{java/lang/Class)}]
  \text{.} 
    \end{align*} 
    }
\item{\textit{Method invocation:}}
Reflection method invocation in byte code is achieved by invoking the API $\syn{java/lang/reflect/Method/invoke}$   via $\syn{invoke}\mhyphen\syn{virtual}$ on a method object, which can be obtained from the first argument.
 The second argument is the receiver object that the method will be invoked on.\footnote{It will be $\syn{null}$ if the method to be invoked is a static method. Its transition rule can be easily adapted from the rule presented.} The third argument is an array of arguments.

{\fontsize{10.5pt}{10.5pt}\selectfont
 $ \overbrace{
 (\sembr{( \syn{invoke} \mhyphen\syn{virtual} \; \syn{java/lang/reflect/Method/invoke} \;(\aexpr_0 \;\aexpr_1\; \aexpr_2)\; \mathit{types_{\mathit{args}}}) :  \vec{s}}, \afp, \astore,  \aka)}^{\aconf}$
 }{\fontsize{10.5pt}{10.5pt}\selectfont
  ~~~$ \aTo 
  \widehat{\mathit{applyMethod}}(  m, \vec{\aexpr}, \afp, \astore, \aka).$
    }
    
    Like the general rule for function call, we have to resolve the method $m$, and then by using $\widehat{\mathit{applyMethod}}$ we can get successor states.

\end{itemize}